# Graphene nanoribbons for quantum electronics

Haomin Wang[1,2,3,#,*], Hui Shan Wang[1,2,3,#], Chuanxu Ma[4,#], Lingxiu Chen[1,2,3,5], Chengxin Jiang[1,3,6], Chen Chen[1,2,3], Xiaoming Xie[1,2,3,6,*], An-Ping Li[7,*], Xinran Wang[8,*]

1 State Key Laboratory of Functional Materials for Informatics, Shanghai Institute of Microsystem and Information Technology, Chinese Academy of Sciences, Shanghai, P. R. China
2 Center of Materials Science and Optoelectronics Engineering, University of Chinese Academy of Sciences, Beijing, P. R. China
3 CAS Center for Excellence in Superconducting Electronics (CENSE), Shanghai, P. R. China
4 Hefei National Laboratory for Physical Sciences at the Microscale and Synergetic Innovation Center of Quantum Information & Quantum Physics, University of Science and Technology of China, Hefei, Anhui, P. R. China
5 School of Materials Science and Physics, China University of Mining and Technology, Xuzhou, Jiangsu, P. R. China
6 School of Physical Science and Technology, ShanghaiTech University, Shanghai, P. R. China
7 Center for Nanophase Materials Sciences, Oak Ridge National Laboratory, Oak Ridge, TN, USA.
8 National Laboratory of Solid State Microstructures, School of Electronic Science and Engineering and Collaborative Innovation Center of Advanced Microstructures, Nanjing University, Nanjing, Jiangsu, P. R. China
#These authors contribute equally.
* Corresponding authors:
hmwang@mail.sim.ac.cn, xmxie@mail.sim.ac.cn, apli@ornl.gov, xrwang@nju.edu.cn

**ABSTRACT:** Graphene nanoribbons (GNRs) are a family of one-dimensional (1D) materials carved from graphene lattice. GNRs possess high mobility and current carrying capability, sizable bandgap, and versatile electronic properties tailored by the orientations and open edge structures. These unique properties make GNRs promising candidates for prospective electronics applications including nano-sized field-effect transistors (FETs), spintronic devices, and quantum information processing. To fully exploit the potential of GNRs, fundamental understanding of structure-property relationship, precise control of atomic structures and scalable production are the main challenges. In the last several years, significant progress has been made toward atomically precise bottom-up synthesis of GNRs and heterojunctions that provide an ideal platform for functional molecular devices, as well as successful production of semiconducting GNR arrays on insulating substrates potentially useful for large-scale digital circuits. With further development, GNRs can be envisioned as a competitive candidate material in future quantum information sciences (QIS). In this Perspective, we review recent progress in GNR research and identify key challenges and new directions likely to develop in the near future.

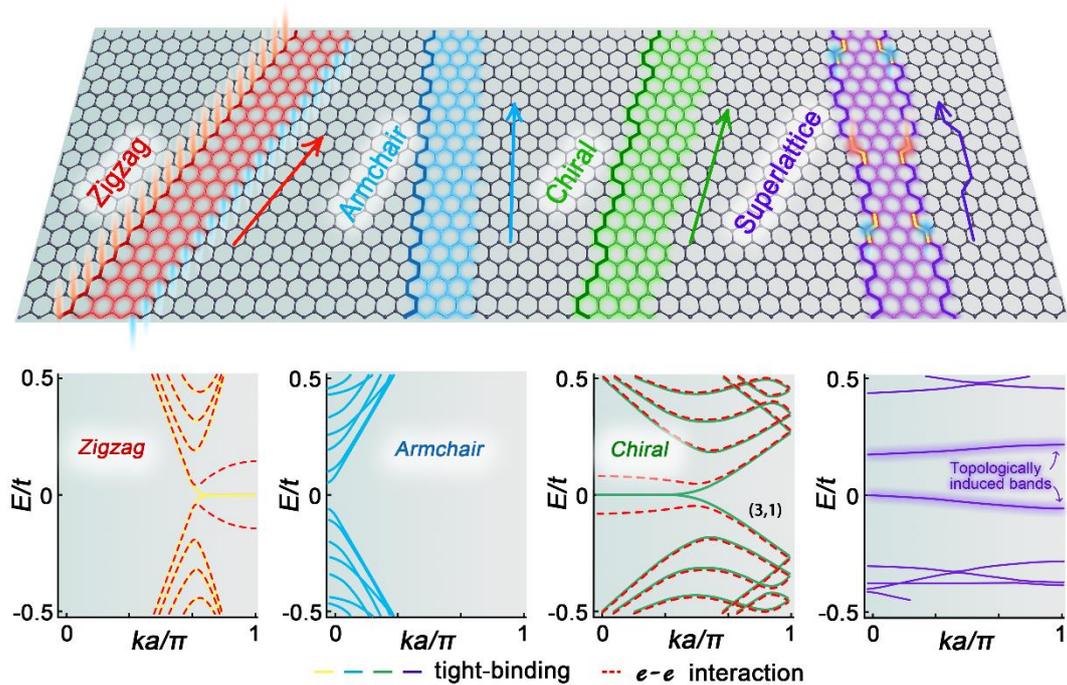

**Box 1 | Electronic structures of GNRs**

According to the edge structure, GNRs can be classified into mainly three categories: zigzag (ZZ), armchair (AC) and chiral GNRs. ZZ and AC edges go through $[11\bar{2}0]$ and $[10\bar{1}0]$ directions of graphene lattice, respectively. Other ordered orientations are chiral, which have a mixture of ZZ and AC edges.

The properties of GNRs highly depend on their edge structures. Nearest-neighbor tight-binding model predicts that all ZZ-edged GNRs (ZGNRs) are metallic[1-3]. The edge states localized at the ZZ edges exhibit ferromagnetic ordering,[2,4,5] and lead to the formation of conducting channels near the Fermi level. Exchange coupling of ferromagnetic states across the ZGNR results in anti-ferromagnetism and bandgap opening when its width is narrow enough.[1,6,7] The half-metallicity is expected by applying an in-plane transverse electric field to lift the spin degeneracy.[6] These magnetic properties render ZGNRs useful in spintronics.[8,9]

AC-edged GNRs (AGNRs) exhibit alternating metallic and semiconducting behavior.[2,4,10,11] Depending on their widths, AGNRs can be classified into three families: $N = 3p$ and $N = 3p + 1$ families ($p$ is an integer and N is the number of carbon-carbon chains across the width) with large $E_g$ inversely proportional to the width[1,3], and the $N = 3p + 2$ family predicted to have very small $E_g$[11,12]. The opening of $E_g$ originates from spatial confinement of electronic wavefunctions like carbon nanotubes. The width scaling also leads to a tradeoff between the mobility and $E_g$ for different applications.[13] AGNRs exhibit relatively large $E_g$ and therefore are suitable for scaled logic transistor,[14-16] while AGNRs with small $E_g$ allow lift of valley degeneracy and thus enable spin valve/filter *via* spin-orbit and electron-phonon coupling.[17,18]

Chiral GNRs have mixed ZZ and AC edges. Segments of ZZ edges present magnetic edge states, where the appearance of both edge-state splitting and magnetic moments depends on the width.[19] A desirable conducting channel is predicted in certain chiral ribbons[20] as backscattering near charge neutrality is forbidden because of the absence of an inversely dispersive band at the same K point of the Brillouin zone.

The GNR superlattice of alternating topologically trivial and non-trivial segments possesses topological states[21,22], which can be exploited for quantum information processing.[23,24] These states offer topologically-protected spin centers, which form an antiferromagnetic spin-½ chain with tunable exchange interactions[21,23,24], making them a promising platform for spintronic devices as well. Thanks to the advances in precise bottom-up assembly, it is now possible to integrate qubits on a single GNR for full-fledged QIS applications.

**Main text**
**1）Introduction**
Over the past half-century, the integration of micro-electronic devices promotes the stupendous development of high-performance computers. However, the feature size of the transistors in integrated circuits has been reduced to nanometers by following Moore's Law[25], and further device miniaturization faces challenges from both fundamental physics and manufacturing technology, such as short-channel effects[26] and the limits of lithography[27]. To achieve higher performance together with smaller device footprint and reduced power consumption, low-dimensional materials are urgently needed to complement bulk semiconductors. Beyond classic computers based on Boolean operation, quantum computing has the potential to revolutionize computation by making certain types of classically intractable problems solvable[28]. The creation of reliable quantum bits (qubits) in solid state systems requires engineering topologically protected states in new materials.[29,30]

In recent years, many low-dimensional materials have been proposed along the "more Moore" and "more than Moore" paths, with the hope that they can offer additional performance or degree of freedom in computing. Among them, GNRs, which can be viewed as quantum confined graphene in one dimension, have attracted enormous interest because of outstanding electronic properties (e.g. sizable $E_g$[2,4], long mean free path (MFP)[31], localized spin[32] and topological edges states[21], see **Box 1**). Furthermore, GNR offers width in the nanometer regime which naturally allows the scaling of device dimensions and ultrahigh density integration. Compared to the other candidates suggested by the International Roadmap for Devices and Systems (IRDS)[33], such as III-V compound semiconductors, Ge, SiGe, two-dimensional (2D) crystals and carbon nanotubes (CNTs), GNRs' tunable $E_g$ and versatile edge states make them highly attractive as building blocks for information processing under both classical and quantum schemes. In realizing topological states and controlling quantum coherence, in particular, GNRs have several potential advantages. *First*, in GNR structures, two main sources of decoherence, spin-orbit coupling and hyperfine interaction, are both minimal. *Second*, the bandwidth of the topological electronic band close to the energy scale of proximity induced spin-orbit coupling can be fine tuned[24]. Controlled periodic coupling of topological boundary states at junctions of GNR creates quasi-1D trivial and non-trivial electronic quantum phases. Zigzag oriented GNR possess spin-polarized edge states.[34] Both of them could serve as key elements for quantum information devices. *Third*, the quantum states in GNR nanostructures, once made, can be readily amenable to integration with a wide family of 2D materials for assembling into multi-qubits architectures and ultimately integrated into systems.

The theoretical investigations on GNRs has initiated since 1996,[1] much earlier than the experimental isolation of monolayer graphene[35]. The theoretical studies investigated the edge and spatial confinement effect using the tight binding calculations.[1,2] More recently, using advanced first-principles calculations considering the orbital hybridizations and electron-electron interactions,[4,6,12,36] the detailed electronic structures of GNRs have been well understood (see **Box 1**). Many device concepts from logic transistors to optoelectronics have been theoretically proposed[37,38] and experimentally realized[39,40]. Recently, topological states in GNRs have been engineered *via* atomically precise synthesis strategy[21]. As such, it is possible to place unpaired electrons at any location designated within a GNR, so long as it contains an interface between segments with different topologies. This approach can be used to create qubits, quantum spin chains and new 1D band structures.

After more than 20 years of research, the potential of GNRs in quantum electronics has now become clear. However, the path forward is not smooth as we are still facing many challenges in scalable production of GNRs with atomic precision, high-throughput characterization, and device integration. In this perspective, we will first review the recent progresses in GNR production, focusing particularly on the ability to precisely engineer the edge structure by bottom-up molecular assembly and to scale up for device integration. Next, we introduce the main challenges and possible solutions in materials, devices and circuitry integration. Finally, we discuss the possible future applications ranging from 3D integration for logic and memory, spintronics to topological quantum information.

## 2）Towards atomic-precise and large-scale production of GNRs

Since 2007, top-down approaches were first adopted for GNR fabrication by lithographic cutting of graphene[41,42], and later on sono-chemical treatment in solution[43], unzipping of carbon nanotubes[44-46] and anisotropic etching technique[47,48]. These approaches obtained sub-10 nm GNRs and proved the opening of sizable $E_g$ by transport and device measurements. An important milestone was the demonstration of high on/off ratio GNR-based FETs in both p-doped[43,49] and n-doped[50] forms, which created world-wide initiatives to study this type of material[51,52]. However, top-down approaches can hardly control the width and especially the edge structure of GNRs at the atomic level, and thus only chemically less defined GNRs can be obtained, stymieing the accurate control of their properties[48,53].

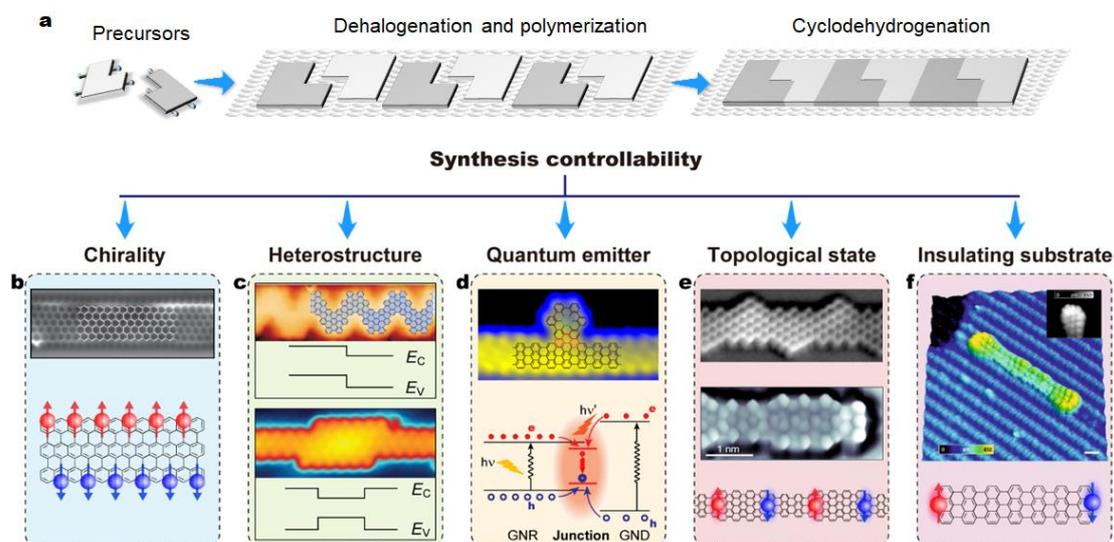

**Fig. 1 | Synthesis update of chemical assembly on catalytic surface. a|** Schematic of the on-surface synthesis of atomically precise GNRs. **b|** 6-ZGNR synthesized on Au(111)[54], of which the ground state of the anti-ferromagnetic spin configuration is schematically illustrated. **c|** GNR heterojunctions and their energy band diagrams[55-57]. **d|** GNR-GND heterojunction formed *via* conjugation between the armchair edges[58]. The schematic shows the formation of quantum-well-like states at the interface, which can act as quantum emitters with light excitation. **e|** GNR superlattices with topologically protected spin states[23,24]. The schematic illustrates the formation of a 1D spin chain in the 7-9 AGNR superlattice. **f |** 7-AGNR synthesized on an insulating rutile (011)-(2×1) surface[59], which can host intrinsic anti-ferromagnetically coupled spins formed at the ends.

Bottom-up assembly uses small molecular precursors as building blocks to create well-defined GNR structures. As illustrated in **Figure 1a**, the on-surface reaction usually follows a two-stage reaction process, i.e., Ullmann polymerization and cyclodehydrogenation. In 2010, Cai *et al.* reported the first bottom-up synthesis of armchair GNRs with width of seven carbon atoms (7-AGNR) by polymerization of 10,10'-dibromo-9,9'-bianthryl molecular precursors, opening up an avenue to the atomically precise production of GNRs[60]. Besides the 7-AGNRs on Au and Ag surfaces, a broad range of GNRs has been realized including AGNRs with 5[61], 7[60,62], 9[24], and 13[63] carbon atoms, ZGNRs with widths of 6 atoms (**Figure 1b**)[54], and chiral-edged GNRs[64,65]. Doped GNRs can be achieved by incorporating heteroatoms in the precursors, such as B[66,67], BN[68], N[69-71], S[72] and O[73], which has proved effective to tune the electronic properties[74]. Combining scanning tunneling microscopy and spectroscopy (STM/STS) characterizations with first-principles calculations, a wide range of $E_g$ (0.1~3 eV) has been demonstrated.

Furthermore, GNR heterojunctions can be rationally designed by welding different precursors (**Figure 1c**)[55-57]. Cai *et al.* and Nguyen *et al.*, respectively demonstrated the formation of a type-II band alignment between the pristine and the N-doped chevron GNRs[55,56] and between the pristine and the fluorenone chevron GNRs[73]. Chen *et al.* reported type-I 7-13 AGNRs heterojunction[57]. Another approach to form GNR heterojunctions is to fuse the parallel GNRs together, as demonstrated by Ma *et al.*[75], giving the 7-14 and 7-14-21 AGNR heterostructures. Similar strategy can be applied to GNR and graphene nanodot (GND) to form a GNR-GND heterojunction with well-defined quantum-well states for narrow-band photoluminescence (PL) (**Figure 1d**)[58].

Engineering topological states through controlled growth of nanostructures enables the formation of topological bands, allowing an increased level of precise control over the electronic structure of such materials. As shown in **Figure 1e**, by mixing precursors and fusing different types of GNR segments[56,57,63], Rizzo *et al.* and Gröning *et al.*, respectively, reported the topologically-protected states at the boundary of the 7-AGNR with staggered edge extensions[24] and at the interfaces between topologically trivial 7-AGNR and nontrivial 9-AGNR segments in the superlattice[23]. This is reminiscent the Su, Schrieffer and Heeger (SSH) chain[24], providing a completely new design space to realize exotic topological states for quantum information processing. Very recently, by inserting a symmetric superlattice of zero-energy modes into otherwise semiconducting GNRs, robust metallic states have been realized in GNRs[76].

So far, most of the bottom-up assembly were demonstrated on noble metals due to their catalytic properties. However, the interaction with metallic substrates strongly affects the electronic and optical properties of the GNRs, and the device application would require the transfer of GNRs off the substrate[12,55-57]. It is therefore highly appealing to grow GNRs directly on insulating substrates. Very recently, Kolmer *et al*. demonstrated the direct synthesis of 7-AGNR on the surface of rutile $TiO_2$ by combining cyclodehydrofluorination and cyclodehydrogenation reactions (**Figure 1f**)[59], where STS characterizations revealed that the magnetic ground state of the synthesized model open-shell GNRs was electronically decoupled from the substrate. The ability of accessing the intrinsic properties of GNRs synthesized by the on-surface approach not only provides more possibilities for evaluating their potential applications, but also offers more convenient platforms for device fabrications in quantum information and quantum computing.

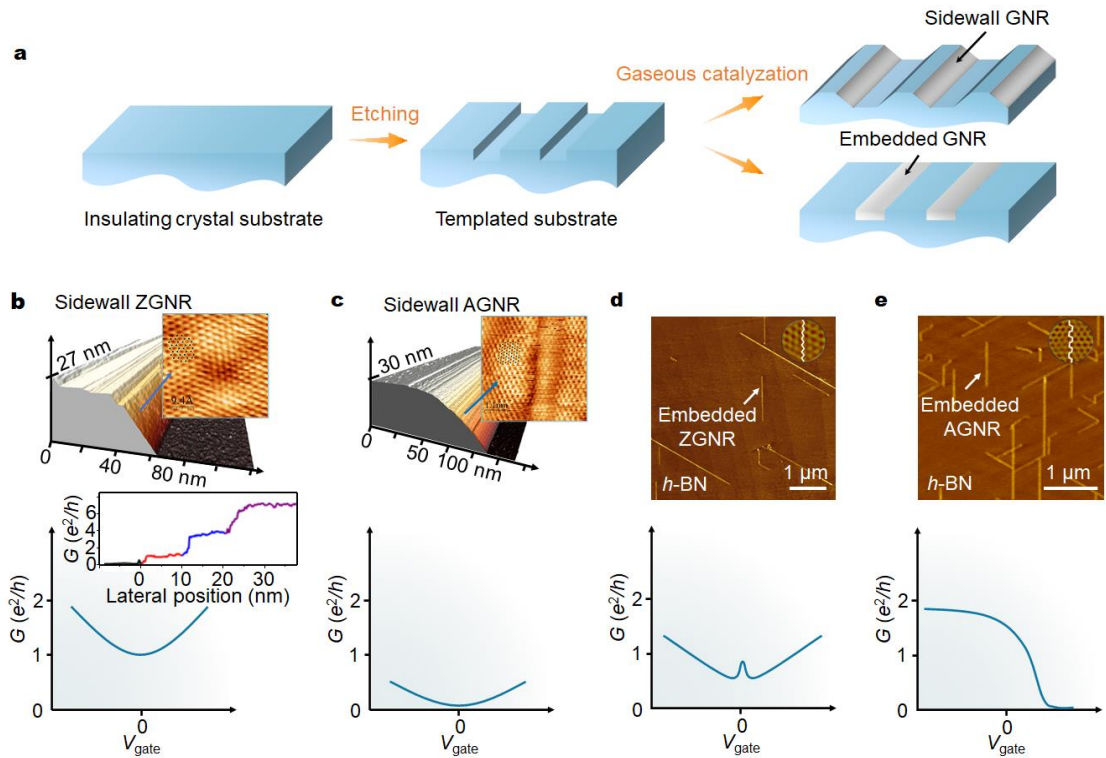

**Fig. 2 | CVD synthesis and epitaxy on insulating surface. a|** Schematic illustration of the growth process of GNRs on SiC and *h*-BN crystals. **b, c|** 3D STM images of a sidewall ZGNR and AGNR on SiC mesa structures oriented along the [1$\bar{1}$00] and [11$\bar{2}$0] direction, respectively.[77] The insets show magnifications. The lower part shows the characteristic transfer curve of a wide sidewall ZGNR or AGNR. The lower inset for sidewall ZGNR shows a measurement result of two-point-probe conductance by moving one of the probes. Transverse movement sequentially reveals the edge channel (red) and body channels (blue, purple) in sidewall ZGNRs.[78] **d, e|** AFM height images of embedded ZGNRs and AGNRs, respectively. The inset circulars show atomic-resolution friction images of *h*-BN. The characteristic transfer curve of a wide embedded ZGNR shows a pronounced conductance peak which is related to the edge states. The FET made of embedded AGNR can be completely switched off even at room temperature[79].

Another important aspect is the scalability. To this end, chemical vapor deposition (CVD) can produce GNRs with well-defined edges and predetermined locations, in a scalable fashion[80-82]. It is also possible to control ZGNRs or AGNRs by engineering the substrate templates. Recently, SiC [83] and *h*-BN[84-86] have been used as template substrates towards this goal. To achieve scalable GNR arrays, sidewall or trench templates need to be first created on the substrate, by lithography or metal catalytic etching. The GNRs are then grown on the templates *via* gaseous CVD (**Figure 2a**). On SiC, the orientation of GNRs is determined by the predefined sidewalls. AGNRs and ZGNRs can be obtained on the [11$\bar{2}$0] and [1$\bar{1}$00] sidewalls of the mesa structures, respectively, showing distinct properties.[77] ZGNRs resemble ballistic conductors, which exhibit quantum resistance of several $e^2/h$ (where $e$ is the elementary charge and $h$ is the Planck constant) and low on-off ratio[31,78,87] (**Figure 2b**). The result is related to the edge geometry and asymmetric termination at opposite edges.[78] In contrast, the sidewall AGNRs appear to be strongly corrugated following the topography of the substrate and show higher resistance. (**Figure 2c**) Surprisingly, even though its width is reduced to just 2-3 nm, the AGNRs on the nano-facets still behave metallic.[88,89] Recently, quasiparticle band gap formation[90,91] in very narrow sidewall AGNRs was observed. These conflicting results indicate that the correlation between ribbon geometry and transport behavior need further investigation.

$h$-BN is an ideal substrate for GNRs due to small lattice mismatch ~1.6%, flat surface free of dangling bonds and charge impurities.[92] On $h$-BN, chirality-controlled GNRs can be grown *via* CVD within pre-defined ZZ or AC trenches from metal nanoparticle etching (**Figure 2d, e**). [79,86,93,94] Narrow embedded ZGNRs (< 7 nm) exhibit $E_g$ greater than 0.4 eV.[79] The gap opening is attributed to a combined influence from $e$–$e$ interactions, uniaxial strain and the stacking order on $h$-BN. Intriguingly, wide ZGNRs always show pronounced conductance peak in the transfer curves, which is temperature independent and persists under magnetic fields. The phenomenon is believed to be related to gap states localized at the ZZ edges.[95] Narrow embedded AGNR is of an $E_g$ of ~530 meV extracted from device characteristics (**Figure 2e**). No obvious conductance peak was observed in AGNRs. It indicates that the opening of $E_g$ is a result of quantum confinement. Importantly, the mobility of the narrow GNRs reaches ~1,500 cm$^2$ V$^{-1}$ s$^{-1}$ presumably due to edge passivation and impurity screening by $h$-BN. In addition, multi-dimensional integration of GNRs with $h$-BN by in-plane covalent bonding[96] and van der Waals stacking[97] brings the heterostructure a merit in chemical/mechanical stability, and may lead to emergent properties that are non-existent from the constituents.

**3）Challenges and future of GNR electronics**

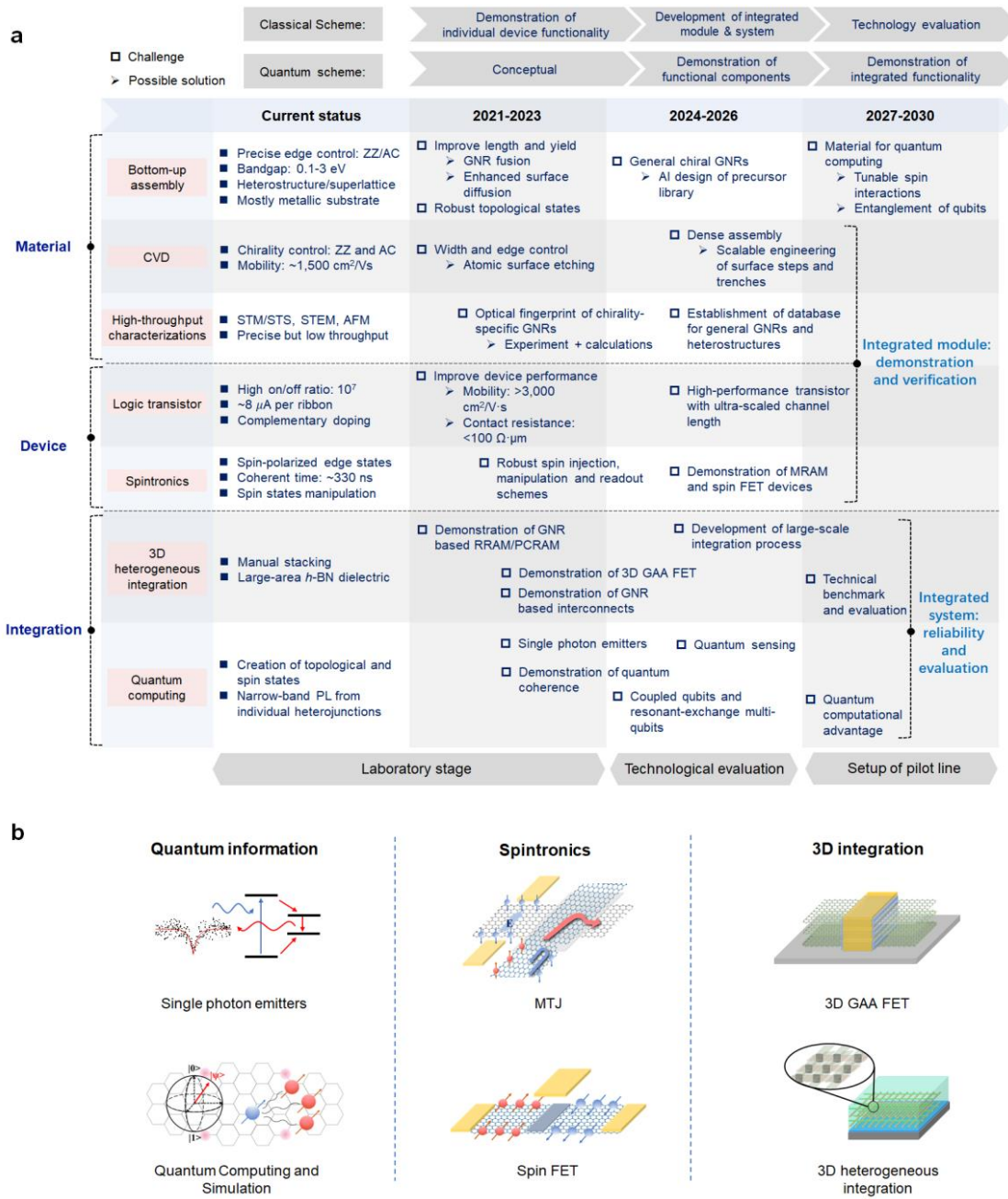

**Fig. 3 | A technology roadmap for GNR electronics. a|** Current status, challenges and their possible solutions in material synthesis, device design and circuit integration. **b|** Emerging applications and device architectures in quantum information, spintronics and 3D integration with the improvement of GNR quality, organization and properties control over quantum coherence, spin and charge.

Despite great progress in fundamental understanding of physical properties and proof-of-concept demonstrations, we still face formidable challenges when attempting to create GNR structures and assemblies that are optimally suited for electronic and quantum applications. In this section, we describe the main challenges in material, device and integration and future directions towards GNR quantum electronics (**Figure 3**).

**Material**

The primary challenge in GNR synthesis is scalable production of narrow GNRs with atomic precision, on arbitrary substrates, and with designed placement/alignment. Up to now, no perfect synthetic method can meet all these criteria. A hybrid approach that combines merits of different synthesis methods may bring the field to a new stage.

Edge structure control of general GNRs remains challenging. So far, GNRs[98,99] only with certain chirality structure have been produced due to a finite library of precursors. However, the synthesis of GNRs in artificially designed structure is still not possible. In long-term, designing a library of precursors by leveraging artificial intelligence (AI)[100] could be feasible solutions.

Reliable formation of devices requires the continuous improvement of GNR length and yield. Limited by surface diffusivity and kinetic factors on substrates, GNRs made from bottom-up assembly have length typically in a range of just tens of nanometers. Engineering the surface of catalytic substrate to enhance surface diffusion or create long smooth steps could benefit the end-fusion of GNRs, therefore extending their length[101,102]. If the strategy can be extended to crystalline insulator, the yield and organization of GNRs produced for devices could be greatly improved.

Approaching these goals needs to develop scalable substrate templates with atomically precise surface. $h$-BN single crystals could be an ideal potential substrate, and recent progresses show that large-area high quality $h$-BN films can be grown by CVD.[92,103] When arbitrary GNRs with robust topological states can be created on insulating substrates, we can envision a quantum system with entanglement of qubits and tunable spin exchange interactions by varying their separation distances that is controlled by the predesigned molecular precursors.

To characterize GNRs in large quantity, development of high-throughput metrology carrying fingerprint of edge specific GNRs is needed. Up to now, STM/STS, STEM and AFM are popular tools to characterize the atomic structures of GNRs. These tools are precise but low throughput. For CNTs, combination of Raman spectroscopy and optical absorption can determine the chirality[104,105]. Similar database could be established for GNRs, but obviously with much higher complexity due to the open edges. Due to the limited chirality available, spectroscopies (such as Raman[106,107] and photoluminescence[58]) need to be performed on these single-chirality GNRs, and first-principles calculations need to be benchmarked against experimental data. The latter can then generate a database of optical fingerprints of general GNRs and heterostructures.

**Logic device**

A major bottleneck for GNR transistors is to achieve competitive performance (mobility, current density, on/off ratio, *etc*.) over CNTs and 2D materials, which share the advantage of low dimensionality. For example, single CNT ballistic transistor can deliver current approaching 25 μA,[108] and monolayer $MoS_2$ can deliver current density of several hundred μA/μm.[109] So far, a saturation current of only ~8 $\mu A$ was realized in n narrow AGNR.[79] Scattering from edge defects significantly degrades the carrier mobility and saturation velocity compared to graphene, especially in narrow GNRs. This leads to a mobility-$E_g$ trade-off. Keeping GNR edges in atomic smoothness and passivating them in *h*-BN lattice can greatly suppress the scattering[79,86]. In addition, encapsulation by *h*-BN can preserve the MFP and mobility due to the atomically flat and charge impurity-free surface of *h*-BN with high optical phonon energy.[110]

Contact resistances ($R_c$) dominates the total resistance in ultra-scaled FETs. Large Schottky barrier from work function mismatch, Fermi level pinning due to gap states and interface contaminations from device fabrication may all contribute to the increase of $R_c$. Unfortunately, little experimental works are dedicated on this topic. Here we borrow the wisdoms from CNT and 2D materials community and propose some approaches to minimize $R_c$. For examples, graphene can be used as interlayer which can prevent the Fermi level pinning effect.[111] Searching for metals with suitable work function to reduce Schottky barrier, formation of covalent bonding of GNR/metal and creating end-bond structure are all alternative technologies that can improve the contact.[112] As a benchmark, modern silicon transistors have $R_c$ less than 100 Ω·μm and typical 2D semiconductor devices have $R_c$ on the order of several thousand Ω·μm.[113]

**Spintronic device**
GNR opens new possibilities for technological co-implementation of spin generation and manipulation on a common circuit platform. From a practical point of view, the spin stiffness predicted for the magnetic edges in GNRs is higher than traditional magnetic materials[114], and thus ZGNRs would possess higher Curie temperatures than room temperature[115]. The spin coherence time in decorated ZGNRs reaches ~330 ns even at room temperature[34]. What is more, GNRs possess long spin correlation length, a low density of state and a tiny featured size. All of the attractive properties are promising for the GNRs to construct building blocks for spintronic circuitry.

Several prototypes of device configurations for controlling the spin transport have been proposed by using the spin-polarized GNRs as building blocks. One of them is magnetic tunnel junction (MTJ), which is a highly desirable component for magneto-resistive random-access memory (MRAM). A typical MTJ is composed of two ZGNRs separated by a thin dielectric layer. One ZGNR selectively filters the spin direction by a transverse electric field[6], while the other ZGNR with defects on one edge generates a pure spin current by quenching magnetic moments and scattering the carriers at the rough edge[116]. Any change in the spin polarization in the two GNRs could result in the variation of tunneling conduction. Spin FET could be another promising prototype to enable an electrical manipulation of spin currents, inspired by the proposal of spin modulator[117]. In the electrical field, the spin–orbit field serves as a medium for electrical control of the spin current[118,119]. In either device concept, however, a short-term target is the ability to inject spins in GNRs and detect their transport. Ferromagnetic resonance of spin pumping[120] is expected to inject spin into GNR without any tunnel barriers. The detection of spin currents could be realized even without the use of magnetic contacts[121]. The realization of spin degree of freedom *via* field effect in GNRs could potentially be used for Boolean logic computing, as well as unconventional schemes such as neuromorphic and probabilistic computing.

**3D heterogeneous integration**

Owing to the nanometer size, GNR is believed as an excellent candidate for scaling down the device footprint under Moore's law. To realize GNR-based device technology, integration of various components into a circuitry is an essential step. Electrostatic control lies at the heart of modern integrated circuits. Introducing the structures of so-called gate-all-around (GAA) FETs can realize a better electrostatic control over channels in GNR-based circuits. In the 3-nm technology node, the silicon complementary metal-oxide-semiconductor (CMOS) technology will switch to stacked nanosheets geometry to suppress leakage. To this end, GNR as a natural nanosheet offers the advantage of ultrathin body compared to bulk semiconductors. Stacking layers of GNRs vertically in 3D will further increase the device density and promise a high $I_{on}$ comparable to ultra-scaled CMOS.

Owing to the ballistic conductivity, GNR can also serve as contact electrodes for cells in phase-change random-access memory (PCRAM)[122] or resistive random-access memory (RRAM)[123]. One-transistor-one-resistor (1T1R) memory cells can be realized by co-integration of GNR-based transistor and nonvolatile memories[122]. A narrow GNR not only can effectively reduce the active volume of phase change material in each cell of PCRAM, but also serve as channel of a FET with high on/off ratio.[124] The 1T1R architecture not only features excellent scalability, but also benefits in-memory computing, which can reduce the amount of data transfer by performing computations inside the memory arrays. Data-centric applications in machine learning and scientific computing benefit the most from reduced memory access.

In future monolithic 3D-architectures, GNRs enable efficient heat dissipation due to their high thermal conductivity of ~5000 W/m·K[125]. It points to a potential use as heat spreader on-chip. Meanwhile, their atomically-thin nature enables smaller tier thickness and ~150% higher packing efficiency than that in traditional integrated circuits[126]. With the decrease of wire pitches, Cu interconnect dramatically suffers from serious scattering at their surfaces and grain boundaries, which leads to increased self-heating and stability issues[127]. A GNR barrier can keep Cu interconnect resistivity near its intrinsic value regardless of wire width scaling.[128] GNR itself exhibits a breakdown current density of ~$10^9$ A/cm² [129], which is 2 orders higher than Cu wires[130]. Engineering multi-layer GNR *via* intercalation doping further increases current-carrying capability[131]. Therefore, GNRs are potentially useful as interconnects in advanced technology nodes.

IRDS predicts that the dimensional scaling of CMOS technology is reaching fundamental limits. Development of large-scale integration process would be fueled by heterogeneous integration of GNRs on silicon platform. From an industrial point of view, research knowledge from technical benchmark and mature evaluation should be developed to electronic design automation (EDA) tools in early pilot lines to access the basis in module integration.

**Quantum computing**

High-quality single-photon emitters (SPEs) and entangled spin centers are at the core for quantum information and quantum computing. SPEs emit one photon at a time with a well-defined frequency and polarization in a deterministic fashion. SPEs play an important role in many leading QIS technologies, with applications in quantum sensing[132], quantum communications[133], and quantum simulation[134]. However, there is still no 'ideal' on-demand, high purity, indistinguishable SPE. Tuning the electronic structures of GNRs by defect engineering provides a unique opportunity for creating SPEs deterministically. Due to the long coherence length and precision structural control in the GNRs, photo-generated excitons can transfer ballistically to localized defect centers to emit high purity photons, as demonstrated by Ma *et al.* in GNR–GND heterojunctions[58]. Obviously, photon statistics analysis needs to be carried to characterize the single-photon purity, generation efficiency, and indistinguishability. These "defect-engineered" states in GNRs enable the creation of a new family of quantum materials, providing an essential avenue for quantum optoelectronics.

Topologically protected electronic states provide a natural route to phase coherence and entanglement, important for quantum information processing and high-efficiency quantum computation[135,136]. However, even for GNRs with predesigned topological states, it is still non-trivial to realize quantum information processing. For example, how to build qubits with long relaxation and coherence time is critical. The spin coherence time in ZGNR decorated with radical molecules that bears electron spins can last an order of microsecond even at room temperature[34]. When the GNRs are coupled to a solid-state device, the interaction between quantum states and the environment is both a handle to encode quantum states and the origin of decoherence. Placing GNRs on a substrate with different wavelength of relevant acoustic phonons and a clean electro-magnetic environment may preserve coherence time for spins. The spin-phonon interaction and environmental noise are two common sources of decoherence. The former is caused by absorption or emission of one phonon in resonance of two electron spin states, while the latter includes random flips of the spin[137,138] and charge instability[139]. Strain tuning[140] to engineer the phononic bandgap could suppress decoherence by weakening the spin-phonon resonant processes.

Another challenge in the development of the GNR-based quantum devices is information manipulation and readout. Semiconducting AGNRs with lifted valley degeneracy allow Heisenberg exchange coupling for spins in these electrostatically defined, neighboring qubits[141]. The spin qubits can be easily manipulated by an external electric field or coupled remote by microwave resonators, whereas the valley qubits can be controlled with an AC electric field without affecting the spin. In addition, simultaneous flipping of the spin and valley states can be achieved in spin–valley qubits using near-range defects, AC electric fields and off-axis magnetic fields. Further understanding of interactions between the quantum states and identification of the environmental and external variables that can suppress decoherence or be utilized to control the quantum behaviors are urgently needed for GNR-based quantum applications.

Implementation of quantum computing necessitates manufacturing capabilities that allow fabrication of arrays of well characterized qubits in a scalable fashion, ability to initialize set of qubits to a known quantum state, and ability to measure/read the state of the qubit at arbitrary time[142]. Local gating[143] is a convenient strategy for reading, manipulating and initializing the qubits in these GNRs. Moreover, GNRs with coupled qubits or resonant-exchange multi-qubits[144,145] serves as building blocks, and arrangement of them provides significant versatility in qubits integration, which facilitates scalable processing into more complex qubits system[146]. To couple the remote multi-qubits, superconducting microwave resonators offer a means to entangle macroscopically separated quantum systems on a chip. The available qubits and their manipulations determine the quantum algorithm, where the operators perform specific functions on one qubit or more[147,148]. In addition, quantum sensing *via* qubits or entangled photons has become a distinct and rapidly growing frontier of QIS. GNR is expected to provide new opportunities for many areas in applied physics and QIS.

**4) Conclusion**
GNRs are promising materials for semiconductor technologies and QIS. In this perspective article, we have discussed the challenges and opportunities towards GNR-based quantum electronics. In particular, we believe that bottom-up synthesis and epitaxy on templated crystalline substrate offer promising solutions to meet materials requirements. Optimization of GNRs enables different transistor building blocks: semiconducting GNRs for the channel, and quasi-metallic ones for interconnects. Meanwhile, methods to fabricate arrays of single-chiral GNRs in well-controlled density and alignment is needed, which will further promote 3D transistor stacking with well-controlled electrostatics. Monolithic 3D integration will harness the promises of GNRs to deliver packing density and energy efficiency beyond bulk semiconductors.

More importantly, novel device concepts need to be further explored, including spintronic and topological devices potentially useful in quantum computing. It is now generally recognized that entanglement and coherence provide the fundamental resource for a new QIS revolution with widespread practical applications in quantum computing, networking, and sensing[135,136]. By precisely positioning defects or junctions to trap electronic states, single and entangled photon emitters can be created, while controlled local geometry in a superlattice can lead to topologically protected spin centers. It is thus possible to create qubits, quantum spin chains, and new 1D band structures in GNR superlattices. This strategy makes GNRs as a promising material system for topological nanodevices and quantum computers with significantly enhanced coherence and entanglement.

Just as achieving control over the physical and electronic properties of semiconductors was instrumental to the rise of the information age, the key to success in the development and application of QIS is to realize materials where rationally designed structures define electrons, spins, and photons into coherent and entangled states to carry energy and information in unique ways. This will allow efficient transport of energy, quantum information processing and computing, ultrasensitive sensors, and low power electronics. To maximize the probability of success for GNR quantum electronics, synergistic efforts by a cross-disciplinary community of synthetic chemists, physicists, materials scientists, and electrical engineers are critical for pushing the GNR research to a new level of mainstream industry technology.

**Acknowledgement**
The work was partially supported by the National Key R&D program (Grant No. 2017YFF0206106), the Strategic Priority Research Program of Chinese Academy of Sciences (Grant No. XDB30000000), National Natural Science Foundation of China (Grant No. 61734003, 61521001, 61927808, 61851401, 91964202, 61861166001, 51861145202, 51772317, 91964102, 12004406, 22002149), the Science and Technology Commission of Shanghai Municipality (Grant No. 20DZ2203600), Leading-edge Technology Program of Jiangsu Natural Science Foundation (Grant No. BK20202005), Collaborative Innovation Center of Solid-State Lighting and Energy-Saving Electronics, the Fundamental Research Funds for the Central Universities, China and Soft Matter Nanofab (SMN180827) of ShanghaiTech University. C.M. acknowledges support from the Chinese Academy of Sciences (CAS). A portion of the work (A.-P. L) was conducted at the Center for Nanophase Materials Sciences (CNMS), which is a DOE Office of Science User Facility, and supported by grant ONR N00014-20-1-2302.

**Author contributions**
X.W. conceived the Perspective article. H.W., A.-P.L., X.X. and X.W. drafted the manuscript with contributions from H.S.W., C. M., L.C., C.J. and C.C. All authors have read, discussed and contributed to the writing of the manuscript.

**Competing interests**

The authors declare no competing interests.